# Outage Probability Analysis of Amplify-and-Forward Cooperative Diversity Relay Networks

Quoc Tuan Nguyen, Vietnam National University Hanoi, Vietnam
D.T. Nguyen, University of Technology Sydney, Australia
Cong Lam Sinh, Vietnam National University Hanoi, Vietnam

***Abstract:*** *In a cooperative diversity relay network, amplify-and-forward (AF) relaying protocol in conjunction with maximum likelihood detection at the destination has proved to be quite competitive to other relaying protocols. The statistical analysis of the fading end-to-end channel gain of the AF relaying protocol, however, is well known as extremely complex, and research work to date have only studied the asymptotic behavior of the outage probability of the network at either very low or very high signal-to-noise ratios (SNR). Most current works circumvent the analytical complexity by first ignoring the effect of AWGN then by using the simple approximated upper bound min(u,v) for the signal-to-noise ratio. The approximated upper bound min(u,v,uvSNR), proposed in this paper, is far better bound than min(u, v) for the entire SNR, which allows us to derive exact analytical expressions to study the effect of AWGN on the network performance. The accuracy of the resulting lower bound for the network's outage probability using the proposed min(u,v,uvSNR) function is very convincing for the entire range of AWGN.*

## 1. INTRODUCTION

It is well known that message coding is no longer effective in improving transmission reliability during deep and slow fading, and *cooperative diversity* transmission has proved to dramatically improve the performance of transmission [1] [2] [3]. In this paper, we deal only with the classical three-terminal relay network using low-complexity cooperative diversity relaying protocols for ease of potential implementation. In these protocols, relay terminals can process the received signal in different ways, the destination terminals can use different types of combining to achieve spatial diversity gain, and source and relay terminals can use repetition code to cope with low-*SNR* transmission under heavy fade conditions. Relaying protocols can be classified broadly into two classes: amplify-and-forward (AF) which uses linear and continuous processing and decode-and-forward (DF) which uses more adaptive non-linear processing. While AF relaying introduces noise amplification, a destination using maximum likelihood (ML) detection can be quite competitive compared to other protocols, particularly when the relay is close to the destination [7]. The less complex cooperative diversity AF relaying is shown to have comparable bit-error-rate (BER) performance to the DF relaying for independent Gaussian channels with path loss [3]. Similarly, in [5] it is shown that the outage capacity of a two-step cooperative system using orthogonal channels is comparable in the three scenarios: no relaying, amplifying relaying and decoding relaying depending on the reliability of the source-to-relay wireless link.

In slowly fading channels, the fading is assumed constant over the length of the message block, i.e. the channel is memory-less in the blockwise-sense, and the strict Shannon capacity of the channel is well defined and achievable. In most practical situations, the channel is non-ergodic and capacity is a random variable, thus no transmission rate can be considered as reliable. In this case, the *outage* probability is defined as the probability that the instantaneous random capacity falls below a given threshold, and capacity versus outage probability is the natural information theoretic performance measure [2]. In order to calculate the outage capacity, because of the complexity of the probabilistic analysis involved, most authors resort to the *max-flow min-cut theorem* [1, 3, 4] to find an upper bound for the outage capacity of the relay channel. An exact performance analysis of the AF protocol is well known to be very mathematically complex and most authors circumvent the challenge by either neglecting the additive noise at the relay or using a *min(u,v)* function as an approximated upper bound for the end-to-end (E2E) signal-to-noise ratio of the network or by both [3] [5] [6] [7] [8]. The focus of this paper, however, is to find more analytically accurate expressions than are currently available for the outage probability of the AF relaying protocol. In many practical applications, including wireless sensor networks, power is limited and *SNR* is usually very low, and the performance of relaying networks in terms of energy efficiency in the low *SNR* regime becomes essential. However, in the low SNR regime, the Shannon capacity is theoretically zero as *SNR* tends to zero and is no longer a useful measure. Therefore in [2] [3] [8], a more appropriate metric called *outage capacity* is defined as the maximal transmission rate for which the outage probability does not exceed a given threshold. When CSI is unavailable to the transmitters, as in most simple implementations in practice, coherent transmission cannot be exploited, hence even full-duplex cooperation, i.e. where terminals can transmit and receive simultaneously, cannot improve the total Shannon capacity of the network. Therefore, in this paper we focus on half duplex operation.

## 2. SYSTEM MODEL AND INFORMATION RATE

### 2.1 System Model and Definition

In cooperative diversity relaying (see Figure 1), the simplest orthogonal operation is the two-phase time-division multiplexing (TDM). In the relay-receive phase at time $n=1, 2,…T/2$, the source transmits the complete message ($N$ symbols) to both the destination and the $M$ relays ($i=1, 2,..., M$),

$$y_{sr_i}[n] = \sqrt{P_s[n]}h_{sr_i}x_s[n]+n_{sr_i}[n]$$
$$y_{sd}[n] = \sqrt{P_s[n]}h_{sd}x_s[n]+n_{sd}[n] \qquad (1)$$

where $x$, $y$, $n$, and $P$ are the *normalized* transmit signal (i.e. $E(|x|^2)=1$) the corresponding received signal, the additive white Gaussian noise (AWGN) of zero mean and variance $\sigma^2$, i.e. $n \sim \mathcal{N}(0, \sigma^2)$ at the receiver, and the transmit power, respectively, and the parameters' double subscript $ij$ is to mean being associated with the channel link from $i$ to $j$. $h_{ij}$ is the channel gain (or loss) from node $i$ to node $j$, being subject to frequency nonselective Rayleigh fading, and is modeled as an independent, circularly symmetric, complex Gaussian random variable with zero mean and variance $\mu_{ij}$. It is well known that the corresponding $|h_{ij}|^2$ is exponentially distributed with mean $\mu_{ij}$. Note that AWGN is associated with each receiver which in turn is associated with a channel link. In the destination there are at least two receivers, hence at least two noise sources.

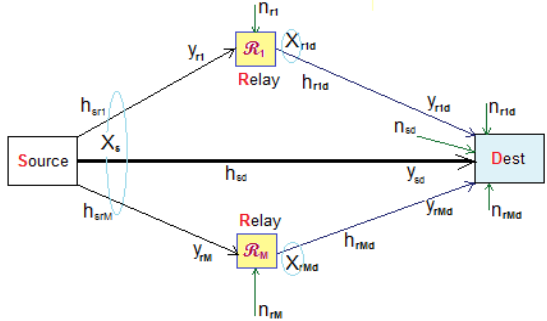

*Figure 1: System model of a cooperative diversity relay network*

In the relay-transmit phase, the relays send their AF signals to the destination. The received signal at the destination is

$$y_{rd}[n+T/2] = \sum_{i=1}^{M}\sqrt{P_{r_i}}h_{r_id}x_{r_i}[n+T/2]+n_{r_id}[n+T/2] \quad (2)$$

which is then combined with the direct signal waiting from the relay-receive phase using maximum ratio combining (MRC). In (2), the transmit signal $x_{ri}$ from the relay is created in two different ways. In the decode-and-forward (DF) relaying mode, the relay detects by fully decoding (or demodulating) the entire codeword it receives from the source, symbol by symbol, then retransmits the signal by recoding (or remodulating) to the destination. While in the amplify-and-forward (AF) relaying mode, the received signal at the relay in (1) is simply amplified by a gain factor $\alpha$ then forwarded to the destination, i.e. $x_{r_i}[n+T/2] = \alpha_{r_i}y_{sr_i}[n]$, then

$$y_{rd}[n+T/2] = \sum_{i=1}^{M}\{h_{r_id}\alpha_{r_i}(\sqrt{P_s[n]}h_{sr_i}x_s[n]+n_{sr_i}[n])+n_{r_id}[n+T/2]\} \quad (3)$$

In order to give the relay the transmit power $P_{ri}$ as in (2) (using an AGC mechanism) the relay gain factor can be calculated by equating the *expected* value of the right hand sides of (2) and (3). The result is

$$\alpha_{r_i} = \sqrt{\frac{P_{ri}}{P_s h_{sr_i}^2 + \sigma_{sr_i}^2}} \quad (4)$$

i.e. in accordance to the $h_{sr}$ channel gain which we assume the relay receiver can estimate accurately.

The destination thus receives ($M$+1) copies of the signal from the source using a maximum ratio combiner (MRC) to obtain the final optimal signal through the maximum likelihood detection.

Below we use the superscript to indicate the relay phase. By rewriting (3), the total received signal at the destination at time $T$ is

$$y_d^{(2)} = \sum_{i=1}^{M}(\alpha_{r_i}\sqrt{P_s^{(1)}}h_{sr_i}h_{r_id}x_s^{(1)}) + \sum_{i=1}^{M}(\alpha_{r_i}h_{r_id}n_{sr_i}^{(1)}) + \sum_{i=1}^{M}n_{r_id}^{(2)}$$

This can be combined with (1) into the matrix below, and for simplicity we put $M$=1,

$$\begin{pmatrix}y_d^{(1)}\\y_d^{(2)}\end{pmatrix} = \begin{pmatrix}\sqrt{P_s^{(1)}}h_{sd} & 0\\ \alpha_r\sqrt{P_s^{(1)}}h_{sr}h_{rd} & 0\end{pmatrix}\begin{pmatrix}x_s^{(1)}\\x_s^{(2)}\end{pmatrix} + \begin{pmatrix}1 & 0 & 0\\ 0 & \alpha_r h_{rd} & 1\end{pmatrix}\begin{pmatrix}n_{sd}^{(1)}\\n_{sr}^{(1)}\\n_{rd}^{(2)}\end{pmatrix} \quad (5)$$

or $\quad \mathbf{Y_d = AX_s + BN}$

### 2.2 Information Rate

The maximum average mutual information between the input and the two outputs, achieved by i.i.d complex Gaussian inputs, of an AF relaying network is

$$I(\mathbf{X}_s;\mathbf{Y}_d|\mathbf{A}) = \frac{1}{1+M}\log_2\{\det(\mathbf{I}_M + \frac{\mathbf{AR_{X_s}A^*}}{\mathbf{BR_NB^*}})\} \quad (6)$$

where $M$ is the number of relays; and the covariance matrices of the input signal and the noise are, respectively, $\mathbf{R_X}=E\{\mathbf{X_sX_s^*}\}=P_s\mathbf{I}$ assuming $P_s^{(1)} = P_s^{(2)} = P_s$ over a period of $T/2$ each phase, and all noise sources are i.i.d with variance $\sigma^2=N_0$, i.e. $\mathbf{R_N} = E\{\mathbf{NN^*}\} = N_0\mathbf{I}$.

$$\mathbf{AR_{X_s}A^*} = \begin{pmatrix}P_s^{(1)}|h_{sd}|^2 & \alpha_r P_s^{(1)}h_{sd}h_{sr}^*h_{rd}^*\\ \alpha_r P_s^{(1)}h_{sd}^*h_{sr}h_{rd} & \alpha_r^2 P_s^{(1)}|h_{sr}|^2|h_{rd}|^2\end{pmatrix}$$

$$\mathbf{BR_NB^*} = \begin{pmatrix}N_0 & 0\\ 0 & N_0 + |\alpha_r h_{rd}|^2 N_0\end{pmatrix}$$

Then

$$det\left(I_M + \frac{AR_{X_s}A^*}{BR_NB^*}\right) = 1 + \frac{P_s^{(1)}|h_{sd}|^2}{N_0} + \frac{P_s^{(1)}\alpha_r^2|h_{sr}|^2|h_{rd}|^2}{(1+\alpha_r^2|h_{rd}|^2)N_0}$$

With $\alpha_r$ in (4), the information rate in (6) using only one relay becomes

$$I_{AF} = \frac{1}{2}log_2\left(1 + |h_{sd}|^2 SNR + \frac{|h_{sr}|^2|h_{rd}|^2}{|h_{sr}|^2+|h_{rd}|^2+\frac{1}{SNR}}SNR\right) \quad (7)$$

In which we denote in *italic* $SNR = P_s/N_0$.

Let the instantaneous end-to-end fading channel gain of the AF cooperative diversity relay network, be

$$|h_{AF}|^2 = |h_{sd}|^2 + \frac{|h_{sr}|^2|h_{rd}|^2}{|h_{sr}|^2+|h_{rd}|^2+1/SNR} \quad (8)$$

We define the *instantaneous* signal-to-noise ratio (SNR) in the received signal as

$$\gamma_{ij} = \frac{|h_{ij}|^2 P_i}{\sigma_{ij}^2} = |h_{ij}|^2 \gamma_{ijAWGN} = |h_{ij}|^2 SNR \quad (9)$$

For convenience, and to be consistent with many papers on the subject, in this paper we have simply used $SNR$ to mean $\gamma_{AWGN}$, the SNR of the unfaded AWGN channel. Under Rayleigh fading, $\gamma_{ij}$ in (9) is an independent *exponential* random variable with expected (average) value

$$\bar{\gamma}_{ij} = \frac{\mu_{ij}^2 P}{N_0} = \mu_{ij}^2 \gamma_{ijAWGN} \quad (10)$$

The maximum instantaneous mutual information of an AF relaying network, from (7) and (8), is

$$I_{AF} = \frac{1}{2}log_2(1 + |h_{AF}|^2 SNR) \quad (11)$$

### 3. E2E SNR AND CHANNEL GAIN

#### 3.1 Exact formula for end-to-end SNR

From the second row of (5) for a single two-hop relay case

$$y_{rd} = h_{sr}h_{rd}\alpha_r\sqrt{P_s}x_s + n_R$$

where $n_R = h_{rd}\alpha_r n_{sr} + n_{rd}$

The instantaneous *SNR* at the destination of the *relayed* signal can be obtained using $\alpha_r$ from (4), as

$$\gamma_R = \frac{|\alpha_r h_{sr} h_{rd}|^2 P_s}{|\alpha_r h_{rd}|^2 \sigma_{sr}^2 + \sigma_{rd}^2} = \frac{\gamma_{sr}\gamma_{rd}}{\gamma_{sr} + \gamma_{rd} + 1} \quad (12)$$

where $\gamma_{sr} = \frac{|h_{sr}|^2 P_s}{\sigma_{sr}^2}$, $\gamma_{rd} = \frac{|h_{rd}|^2 P_r}{\sigma_{rd}^2}$.

The total SNR of the MRC output signal at the destination is

$$\gamma_{AF} = \gamma_{sd} + \frac{\gamma_{sr}\gamma_{rd}}{\gamma_{sr} + \gamma_{rd} + 1} \quad (13)$$

where $\gamma_{sd} = \frac{|h_{sd}|^2 P_s}{\sigma_{sd}^2}$ is the SNR at the destination receiver of the direct link from the source.

For *M*-relay case, the total SNR of the MRC output signal at the destination is the sum of all SNRs of all input signals to the combiner, i.e. $\gamma_{AF} = \gamma_{sd} + \sum_{i=1}^{M} \frac{\gamma_{sr_i}\gamma_{r_id}}{\gamma_{sr_i} + \gamma_{r_id} + 1}$

### 3.2 Approximated upper bound of end-to-end SNR

Since $I_{AF}$ in (11) is a continuous function, the outage probability of the network is defined simply as

$$P_{AF}^{out}(\mu_{th}) = \Pr(|h_{AF}|^2 < \mu_{th}) \quad (14)$$

An exact expression for the statistical distribution of $|h_{AF}|^2$ in (8) is well known to be very difficult to derive, and hence an exact close form solution for the outage probability in (14) is not currently available in the published literature. Most researchers to date prefer to use the following approximated upper bound for the SNR of the two-hop relay channel for *medium and high* SNRs [3] [5] [6] [7] [8],

$$\gamma_R = \frac{\gamma_{sr}\gamma_{rd}}{\gamma_{sr} + \gamma_{rd} + 1} \leq \min\{\gamma_{sr}, \gamma_{rd}\} \quad (15a)$$

or equivalently

$$|h_R|^2 = \frac{|h_{sr}|^2|h_{rd}|^2}{|h_{sr}|^2 + |h_{rd}|^2 + 1/SNR} \leq \min\{|h_{sr}|^2, |h_{rd}|^2\} \quad (15b)$$

The 'bottle neck' approximation in (15a) and (15b) is intuitively arrived at, using the analogy to a series connection of two electrical conductances. It is mathematically very tractable because it facilitates the calculation of the statistical distribution of the end-to-end fading channel gain in (8), hence the outage probability in (14) under various fading conditions. However, under low SNRs and deep fading conditions, the above approximation is quite inaccurate as demonstrated by our work below. If we rewrite (15a) and (15b), respectively as,

$$\gamma_R = \frac{1}{\frac{1}{\gamma_{sr}} + \frac{1}{\gamma_{rd}} + \frac{1}{\gamma_{sr}\gamma_{rd}}}, \quad (16a)$$

and

$$|h_R|^2 = \frac{1}{\frac{1}{|h_{sr}|^2} + \frac{1}{|h_{rd}|^2} + \frac{1}{|h_{sr}|^2|h_{rd}|^2 SNR}} \quad (16b)$$

then for *all* SNRs, we propose the following approximation

$$\gamma_R \leq \min\{\gamma_{sr}, \gamma_{rd}, \gamma_{sr}\gamma_{rd}\} \quad (17a)$$

or equivalently

$$|h_R|^2 \leq \min\{|h_{sr}|^2, |h_{rd}|^2, |h_{sr}|^2|h_{rd}|^2 SNR\} \quad (17b)$$

From the graphs in Figure 2, we can see that when the channel gains are small during deep fading, i.e. small $\mu_u$ and $\mu_v$, the accuracy of the approximation in (17b) is excellent in *both* large and small SNR regimes. Also when the relay position is nearer to one end (large disparity between $\mu_u$ and $\mu_v$) the proposed approximation is better than when the relay is at near equidistance from the ends. This fact can be explained by examining the relative magnitude of the terms in the denominator of (16b) for the two relay locations.

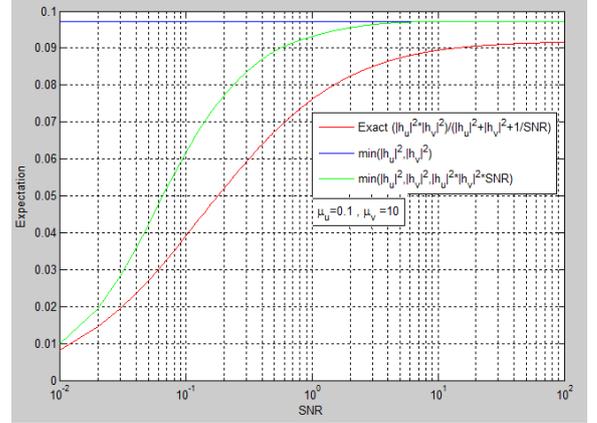

**Figure 2:** *Expected value of the fading gain of the two-hop relaying channel using the exact expression shown in red, using the current upper bound approximation in (15b) shown in blue, and using the proposed upper bound approximation in (17b) shown in green.*

## 4. OUTAGE PROBABILITY ANALYSIS BASED ON E2E SNR

### 4.1 Definition of Outage

The outage probability of the information rate for a given threshold $R_{th}$ is defined as:

$$P_{out}(R_{th}) = P(I < R_{th}) = 1 - P(I \geq R_{th})$$

Or equivalently, using fading channel gain $|h|^2$

$$P_{|h|^2}^{out}(SNR, R_{th}) = \Pr(|h|^2 < \mu_{th} = \frac{2^{(M+1)R_{th}} - 1}{SNR}) \quad (18)$$

In this section, we present accurate expressions for the cumulative distribution function (cdf) of the fading channel gain of a cooperative diversity relay network using an amplify-and-forward relaying protocol. Current research works only report the asymptotic behaviour of the cdf of various relaying protocols at either high or low *SNRs*. The cdf function $F(\mu)$ is used to calculate the outage probability, $P^{out}$, in (18).

There are two asymptotic scenarios associated with $\mu_{th} \to 0$ in (18): one is for very large *SNR* and a given outage threshold $R_{th}$, and the other is for *both SNR* and $R_{th}$ being very small concurrently. In the latter case, $R_{th}$ is quivalent to the ϵ-outagse capacity $C_\epsilon$ [2] [8]. Therefore the limits of the cdf as $\mu_{th} \to 0$ for both asymptotic cases are identical. This is one of the main advantages of our analysis.

### 4.2 Using approximate upper bound $min.(|h_{sr}|^2, |h_{rd}|^2)$

Since the two channel fading gains are independent of each other,

$$F_{|h_R|^2}(\mu) = 1 - Pr(|h_{sr}|^2 \geq \mu)Pr(|h_{rd}|^2 \geq \mu) \quad (19)$$

and (19) can be obtained from (A7) for Rayleigh fading to be

$$F_{|h_R|^2}(\mu) = 1 - exp\left\{-\frac{\mu}{M_r}\right\} \quad (20)$$

i.e. an exponential random variable with mean $M_r$,

where $M_r = \left\{\frac{1}{\mu_{sr}} + \frac{1}{\mu_{rd}}\right\}^{-1}$

The end-to-end fading gain can be *approximated* by its *upper* bound as
$$|h_{AF}|^2 = |h_{sd}|^2 + min\{|h_{sr}|^2, |h_{rd}|^2\} \quad (21)$$
Thus the cdf of $|h_{AF}|^2$ in (21) can be obtained from (A3) as the convolution of (A1) and (A7), and it is
$$F_{|h_{AF}|^2}(\mu) = \frac{1}{(1/\mu_{sd} - 1/M_r)} \cdot$$
$$\left[\frac{1}{\mu_{sd}}\left(1 - e^{-\mu/M_r}\right) - \frac{1}{M_r}\left(1 - e^{-\mu/\mu_{sd}}\right)\right] \quad (22)$$
Therefore $P_{AF}^{out}(SNR, R_{th}) = F_{|h_{AF}|^2}(\mu_{th})$ (23)

### 4.3 Using approximate upper bound $min.(|h_{rd}|^2, |h_{sr}|^2, |h_{sr}|^2|h_{rd}|^2 SNR)$

Using (A7) and (A11), the more accurate approximated upper bound in (17b) readily give
$$F_{|h_R|^2}(\mu) = 1 - Pr(|h_{sr}|^2 > \mu)$$
$$Pr(|h_{rd}|^2 > \mu)Pr(|h_{sr}|^2|h_{rd}|^2 > \mu/SNR)$$
$$= 1 - exp\left\{-\mu/M_r\right\} 2\sqrt{\frac{\mu/SNR}{\mu_{sr}\mu_{rd}}} K_1\left(2\sqrt{\frac{\mu/SNR}{\mu_{sr}\mu_{rd}}}\right) \quad (24)$$
And since
$$|h_{AF}|^2 = |h_{sd}|^2 + min\{|h_{sr}|^2, |h_{rd}|^2, |h_{sr}|^2|h_{rd}|^2 SNR\}$$
the convolution relation of sum of two random variables gives
$$F_{|h_{AF}|^2}(\mu) = \int_0^\mu f_{|h_{sd}|^2}(x) F_{|h_R|^2}(\mu - x) dx$$
Then $F_{|h_{AF}|^2}(\mu) = \int_0^\mu \frac{1}{\mu_{sd}} exp\left(-\frac{x}{\mu_{sd}}\right) \cdot$
$$\left\{1 - exp\left(-\frac{(\mu-x)}{M_r}\right) 2\sqrt{\frac{(\mu-x)/SNR}{\mu_{sr}\mu_{rd}}} K_1\left(2\sqrt{\frac{(\mu-x)/SNR}{\mu_{sr}\mu_{rd}}}\right)\right\} dx$$
Let $y=\mu-x$
$$F_{|h_{AF}|^2}(\mu) = 1 - exp\left(-\frac{\mu}{\mu_{sd}}\right) - \frac{1}{\mu_{sd}} exp\left(-\frac{\mu}{\mu_{sd}}\right) \cdot$$
$$\int_0^\mu exp\left[-y\left(\frac{1}{M_r} - \frac{1}{\mu_{sd}}\right)\right] 2\sqrt{\frac{y/SNR}{\mu_{sr}\mu_{rd}}} K_1\left(2\sqrt{\frac{y/SNR}{\mu_{sr}\mu_{rd}}}\right) dy \quad (25)$$
Finally the outage probability can be calculated as in (23).

### 4.4 Using cut-set bound

Using the *max-flow min-cut theorem* [1] [3] yields the *upper bound* of the capacity of a *general full duplex* relaying system with multiple input and multiple output (MIMO), in which transmit and receive signals occur concurrently in the same time slot. It is the upper bound for capacity because this is when both the *broadcast channel* (BC) and the multiple access channel (MAC) channels are in full diversity connection. The AF relaying is a general relay channel, therefore we use [1, Theorem 3]
$$C^+ = \max_{f(X_1, X_2)} \min \left\{I(X_1; (Y_2, Y_3)|X_2), I((X_1, X_2); Y_3)\right\} \quad (26)$$
Thus, the upper bound for capacity, in the case of no correlation between $X_1$ and $X_2$ and equal transmit power from the source and the relay, is
$$C^+ = \min\left\{\frac{1}{2}\log(1+(\gamma_{sd}+\gamma_{sr})), \frac{1}{2}\log(1+(\gamma_{sd}+\gamma_{rd}))\right\} \quad (27)$$
Equivalently from (27), the cut-set-bound of the end-to-end network gain is
$$|h_{CSB}|^2 = min\{(|h_{sd}|^2 + |h_{sr}|^2), (|h_{sd}|^2 + |h_{rd}|^2)\} \quad (28)$$
The corresponding *lower* bound of the outage probability under exponential fading condition is
$$P_{CSB}^{out}(\mu_{th}) = 1 - Pr[(|h_{sd}|^2+|h_{sr}|^2)>\mu_{th}] \cdot Pr[(|h_{sd}|^2+|h_{rd}|^2)>\mu_{th}]$$
$$= 1 - \frac{1}{\mu_{sd}-\mu_{sr}}\left\{\mu_{sd}e^{-\frac{\mu_{th}}{\mu_{sd}}} - \mu_{sr}e^{-\frac{\mu_{th}}{\mu_{sr}}}\right\} \cdot$$
$$\frac{1}{\mu_{sd}-\mu_{rd}}\left\{\mu_{sd}e^{-\frac{\mu_{th}}{\mu_{sd}}} - \mu_{rd}e^{-\frac{\mu_{th}}{\mu_{rd}}}\right\} \quad (29)$$
The result in (29) can be obtained by using (A3) and (A7) of the Appendix. Since
$$E[|h_{sd}|^2 + min(|h_{sr}|^2, |h_{rd}|^2)]$$
$$\geq E[min(|h_{sd}|^2 + |h_{sr}|^2, |h_{sd}|^2 + |h_{rd}|^2)] \quad (30)$$
$$\geq E[|h_{sd}|^2 + min(|h_{sr}|^2, |h_{rd}|^2, |h_{sr}|^2|h_{rd}|^2 SNR)]$$
The approximation that has been most used in the literature, i.e. using $min(u,v)$ is the worst of all upper bounds. In Figures 2 and 3, we have not plotted the results corresponding to the cut-set bound because it can be easily seen from (30) that this bound is almost the same as the $min(u,v)$ bound.

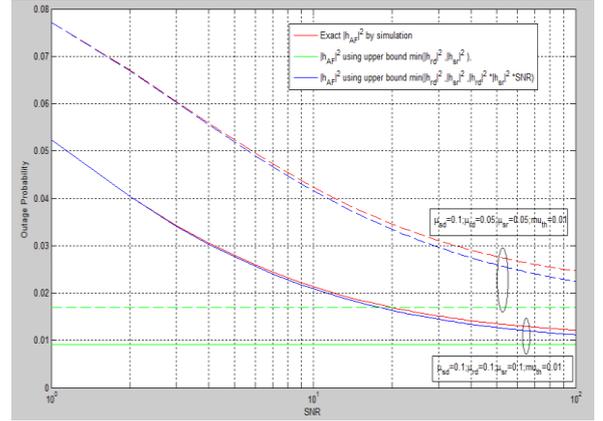

**Figure 3:** *Effect of signal-to-AWGN noise ratio on the outage probability of a cooperative diversity relaying network using various approximations and bounds.*

## 4. CONCLUSIONS AND DISCUSSION

The statistical analysis of the instantaneous fading end-to-end signal-to-noise ratio or its equivalent channel gain of the AF relaying protocol is well known as extremely complex, and research works to date only study the asymptotic behavior of the outage probability of the network at *either* very low *or* very high signal-to-noise ratios (SNR). In this paper, we have made a successful step towards a more accurate analysis than is currently available for the complete range of *SNR*. The outage probability of the cooperative diversity relay network using AF relaying protocol has been calculated as a function of the outage threshold, $\mu_{th}$, of the end-to-end fading channel gain. The advantage of this threshold parameter is that both asymptotic scenarios, large SNR-finite rate and low SNR-low rate, may be studied by letting $\mu_{th}$ tending to zero.

Most current works circumvent the analytical complexity by first ignoring the effect of AWGN then by using the simple approximated upper bound $min(u,v)$ for the signal-to-noise ratio in (15a) or equivalently the fading channel gain in (15b). We can see from Figure 2 that our proposed approximated upper bound $min(u,v,uvSNR)$ is far better bound than $min(u,v)$ for the entire *SNR*, which allows us to study the effect of AWGN on the network performance, in particular at low SNRs in many battery-powered cognitive radio and remote wireless sensor networks. In Figure 3, the superior accuracy of the resulting lower bound for the network's outage probability using the proposed $min(u,v,uvSNR)$ function is very convincing for the entire range of AWGN.

The paper, indeed, has made a significant step towards an exact solution for the outage probability of the cooperative AF

relaying protocol, but the challenge of the exact solution remains finding the closed form for the integration in (25).

## APPENDIX

### 1. Distribution of a single exponential random variable

Let $u$ be an exponential r.v. with mean $\mu_u$, then
$$f_U(u) = \frac{1}{\mu_u} e^{-u/\mu_u} \quad F_U(u) = 1 - e^{-u/\mu_u} \quad (A1)$$
Then $\quad lim_{\mu \to 0} \left\{ \frac{F_U(\mu)}{\mu} \right\} = \frac{1}{\mu_u}$ (A2)
by using the approximation $e^{-x} = 1 - x$ for $x \ll 1$

### 2. Distribution of sum of two independent exponential random variables

Let $s = u + v$, where $u, v$ are two independent exponential r.v.'s with mean $\mu_u$ and $\mu_v$, respectively, then from the convolution theorem
$$f_s(\mu) = (f_U \oplus f_V)_\mu$$
$$= \frac{1}{\mu_u \mu_v} \int_0^\mu e^{-x/\mu_u} e^{-(\mu-x)/\mu_v} dx = \frac{e^{-\mu/\mu_v} - e^{-\mu/\mu_u}}{\mu_v - \mu_u}$$
Hence
$$F_s(\mu) = \mu \int_0^\mu f_s(x) dx$$
$$= \frac{1}{\mu_v - \mu_u} \{ \mu_v (1 - e^{-\mu/\mu_v}) - \mu_u (1 - e^{-\mu/\mu_u}) \} \quad (A3)$$
By using the approximation, $e^{-x} \approx 1 - x + x^2/2$ we obtain
$$lim_{\mu \to 0} \left\{ \frac{F_s(\mu)}{\mu^2} \right\} = \frac{1}{2\mu_v \mu_u} \quad (A4)$$
which can be generalized to the case of $s = \sum_{i=0}^{K} u_i$
$$lim_{\mu \to 0} \left\{ \frac{F_s(\mu)}{\mu^{K+1}} \right\} = \frac{1}{(K+1)!} \prod_{i=0}^{K} \frac{1}{\mu_i} \quad (A5)$$
If $z = u + v + c$, where $c$ is a constant, then
$$f_Z(\mu) = f_S(\mu + c) \quad (A6)$$

### 3. Distribution of the Minimum independent exponential random variables

Let $m = \min(u, v)$ where $u, v$ are independent exponential random variables with mean $\mu_u$ and $\mu_v$, respectively. For $m > \mu$, all terms in $\min(u,v)$ should be $> \mu$. Therefore the complementary cdf of $m$ is
$$F_M(\mu) = 1 - F_M(m \geq \mu) = 1 - P(u \geq \mu, v \geq \mu)$$
Since $u$ and $v$ are independent of each other, we have $F_M(\mu) = 1 - P(u \geq \mu) P(v \geq \mu)$. For exponential distributions, it is easy to obtain
$$F_M(\mu) = 1 - exp\left\{ -\mu \left( \frac{1}{\mu_u} + \frac{1}{\mu_v} \right) \right\} \quad (A7)$$
i.e. $m$ is an exponential r.v. having mean $\mu_m$ which is
$$\frac{1}{\mu_m} = \frac{1}{\mu_u} + \frac{1}{\mu_v}$$
Also from (A7), $lim_{\mu \to 0} \left\{ \frac{F_M(\mu)}{\mu} \right\} = \frac{1}{\mu_u} + \frac{1}{\mu_v}$ (A8)
(A8) can be generalized to the case of $K$ exponentials,
$$\frac{1}{\mu_w} = \sum_i^K \frac{1}{\mu_i} \quad (A9)$$
Note: The distribution of $\max(u,v)$ is *not* an exponential r.v.

### 4. Distribution of Product of independent exponential random variables

Let $p = u.v$, where $u > 0$, $v > 0$ are two independent exponential r.v's of mean $\mu_u$ and $\mu_v$, respectively, then by using the Jacobian transform method, we obtain
$$f_p(p) = \int_0^\infty \frac{1}{z} f_U\left(\frac{p}{z}\right) f_V(z) dz = \frac{1}{\mu_u \mu_v} \int_0^\infty \frac{1}{z} e^{-\frac{p}{\mu_u z}} e^{-\frac{z}{\mu_v}} dz$$
Note that dimension of $p$ is $\mu^2$.

From [9, §3.471.9 p.368] with $\nu = 0$, $\beta = p/\mu_u$, $\gamma = 1/\mu_v$ )

$$f_p(p) = \frac{1}{\mu_u \mu_v} K_0\left( 2\sqrt{\frac{p}{\mu_u \mu_v}} \right) \quad (A10)$$
where $K_n(x)$ is the modified Bessel function of second kind. Note that the *pdf* of the product of two exponential functions is *not exponential*. The corresponding cdf of $(u.v)$ is
$$F_p(u, v < y) = 1 - \frac{2}{\mu_u \mu_v} \int_y^\infty K_0\left( 2\sqrt{\frac{p}{\mu_u \mu_v}} \right) dp$$
By using [9, §6.592.12, p.691] with $\nu=0$, $\mu=1$, $a = 2 \Big/ \sqrt{\mu_v \mu_u}$
and making a change of variable $p = y.x$, we obtain
$$F_p(u, v < y) = 1 - \frac{2y}{\mu_u \mu_v} \int_1^\infty K_0\left( 2\sqrt{\frac{y}{\mu_u \mu_v}} \sqrt{x} \right) dx$$
$$= 1 - 2\sqrt{\frac{y}{\mu_u \mu_v}} K_1\left( 2\sqrt{\frac{y}{\mu_u \mu_v}} \right) \quad (A11)$$
Using the expansion of $K_1(x)$ for $x < 1$, it can be shown that $x K_1(x) \approx (1 - x^2)$ as $x \to 0$.
$$Lim_{y \to 0} \left\{ \frac{F_p(y)}{y} \right\} = \frac{4}{\mu_v \mu_u} \quad (A12)$$
Again, note that dimension of $y$ is $\mu^2$.

**ACKNOWLEDGEMENT:** This work was supported by a research grant from Project QG 44.10 - TRIGB at UET, Vietnam National University Hanoi.